%% file: umap2025-styles.tex
\documentclass[sigconf]{acmart}

\usepackage{multirow} 
\usepackage{tikz}

\usepackage{array}

\AtBeginDocument{%
  \providecommand\BibTeX{{%
    \normalfont B\kern-0.5em{\scshape i\kern-0.25em b}\kern-0.8em\TeX}}}

\copyrightyear{2025}
\acmYear{2025}
\setcopyright{acmlicensed}\acmConference[UMAP '25]{33rd ACM Conference on User Modeling, Adaptation and Personalization}{June 16--19, 2025}{New York City, NY, USA}
\acmBooktitle{33rd ACM Conference on User Modeling, Adaptation and Personalization (UMAP '25), June 16--19, 2025, New York City, NY, USA}
\acmDOI{10.1145/3699682.3728353}
\acmISBN{979-8-4007-1313-2/2025/06}

\usepackage{booktabs} 
\usepackage{color}
\usepackage{xspace}
\usepackage{framed}

\definecolor{lavendermist}{rgb}{0.9, 0.9, 0.98}
\definecolor{aliceblue}{rgb}{0.94, 0.97, 1.0}
\definecolor{antiquewhite}{rgb}{0.94, 1, 1}

\newcommand{\inv}{\textbf{\texttt{HighInv}}\xspace}
\newcommand{\con}{\textbf{\texttt{HighCon}}\xspace}
\newcommand{\agency}{\textbf{\texttt{Agency}}\xspace}

\begin{document}

\fancyhead{}
\title{Should We Tailor the Talk? Understanding the Impact of Conversational Styles on Preference Elicitation in Conversational Recommender Systems}

\author{Ivica Kostric}
\affiliation{%
  \institution{University of Stavanger}
  \city{Stavanger}
  \country{Norway}
}
\email{ivica.kostric@uis.no}

\author{Krisztian Balog}
\affiliation{%
  \institution{University of Stavanger}
  \city{Stavanger}
  \country{Norway}
}
\email{krisztian.balog@uis.no}

\author{Ujwal Gadiraju}
\affiliation{%
  \institution{Delft University of Technology}
  \city{Delft}
  \country{Netherlands}
}
\email{u.k.gadiraju@tudelft.nl}

\begin{abstract}
Conversational recommender systems (CRSs) provide users with an interactive means to express preferences and receive real-time personalized recommendations. The success of these systems is heavily influenced by the preference elicitation process. 
While existing research mainly focuses on what questions to ask during preference elicitation, there is a notable gap in understanding what role broader interaction patterns---including tone, pacing, and level of proactiveness---play in supporting users in completing a given task. This study investigates the impact of different conversational styles on preference elicitation, task performance, and user satisfaction with CRSs. 
We conducted a controlled experiment in the context of scientific literature recommendation, contrasting two distinct conversational styles---\textit{high involvement} (fast-paced, direct, and proactive with frequent prompts) and \textit{high considerateness} (polite and accommodating, prioritizing clarity and user comfort)---alongside a flexible experimental condition where users could switch between the two. 
Our results indicate that adapting conversational strategies based on user expertise and allowing flexibility between styles can enhance both user satisfaction and the effectiveness of recommendations in CRSs. Overall, our findings hold important implications for the design of future CRSs. 
\end{abstract}

\ccsdesc[500]{Information systems~Users and interactive retrieval}
\ccsdesc[500]{Human-centered computing~User studies}
\ccsdesc[300]{Human-centered computing~Natural language interfaces}
\ccsdesc[500]{Information systems~Recommender systems}

\keywords{Personalization; Conversational recommender systems; Conversational styles}

\maketitle

\input{umap2025-styles-01}
\input{umap2025-styles-02}
\input{umap2025-styles-03}
\input{umap2025-styles-04}

\input{umap2025-styles-05}

\input{umap2025-styles-06}

\begin{acks}
We thank all the participants for their contribution to this study. An unrestricted gift from Google partially supported this research, alongside the TU Delft AI initiative and the ProtectMe Convergence Flagship project.  
\end{acks}

\balance
\bibliographystyle{ACM-Reference-Format}
\bibliography{umap2025-styles.bib}

\end{document}

%% file: umap2025-styles-01.tex
\section{Introduction}

Conversational recommender systems (CRSs) are interactive platforms that engage users in a dialogue, enabling the expression of preferences and feedback in real-time~\citep{Jannach:2022:ACM}. 
This dynamic interaction not only distinguishes CRSs from traditional recommender systems but also underlies their core advantage: the ability to provide personalized, context-aware recommendations that adapt to the user's current needs and preferences. Consequently, the success of CRSs largely depends on the effectiveness of preference elicitation, as it directly influences the relevance and accuracy of the recommendations~\citep{Pramod:2022:Expert}.

Existing approaches in CRSs emphasize the significance of question-based user preference elicitation, focusing on what to ask users, and how to adapt the recommendations based on their responses~\citep{Gao:2021:AI}. The goal here is to understand and utilize user preferences to improve recommendation accuracy.
However, there is still a limited understanding of how to ask preference elicitation questions within the broader context of a conversation. The interaction between users and CRSs plays a critical role in how effectively the system can support users in completing their tasks, yet there is a notable lack of research focusing on the design choices and conversational styles that lead to successful interactions~\citep{Narducci:2018:AI*IA, Jin:2021:HAIa}.
Recent studies have demonstrated that different conversational styles can significantly influence user engagement, task performance, and perceived cognitive load in online work environments~\citep{Qiu:2020:CHI,qiu2020estimating}.  These findings suggest that conversational styles can indeed have a positive impact on interaction outcomes, underscoring the potential for similar benefits in CRSs.

The main objective of our work in this paper is to investigate how conversational styles influence the preference elicitation process within CRSs, and in turn, affect overall task performance and user satisfaction. We examine the impact of conversational styles through three specific research questions (RQs) within the context of a given CRS:

\begin{itemize}
    \item \textbf{RQ1} What is the impact of various conversational styles on preference elicitation? 
    \item \textbf{RQ2} What is the impact of various conversational styles on task performance?
    \item \textbf{RQ3} What is the impact of various conversational styles on user satisfaction?  
\end{itemize}    

\noindent
To address these research questions, we conducted a controlled experiment within the context of scientific literature recommendations, where study subjects were tasked with finding relevant literature given a topic description. 
This user study involves an interactive agent designed with distinct conversational styles across three experimental conditions: (1) \emph{high involvement}, where the system takes initiative, makes frequent suggestions, and utilizes fast interaction by use of buttons; (2) \emph{high considerateness}, a relatively more passive interaction approach, supporting users in their decisions but not taking initiative; and (3) a \emph{flexible setting} that grants users the agency to switch between styles at will. Each condition is analyzed to assess its impact on the preference elicitation stage, task performance, and user satisfaction in a CRS. 

We recruited 30 master's and PhD students in computer science as participants for our study. They were randomly assigned a set of three tasks from one of two domains: \emph{Computer Science} (expected high familiarity) and \emph{Physics} (expected low familiarity). This division allowed us to examine how domain familiarity influences the effectiveness of different conversational styles in a CRS.

Our main findings indicate that participants with higher domain familiarity prefer the high involvement style for its efficiency and directness, enabling them to express their preferences quickly without what could be perceived as `unnecessary explanations.' In contrast, participants with less familiarity favored the high considerateness style due to its explanatory nature, which helped them understand the choices and make informed decisions. Objective measures of task performance confirm the benefits of the high considerateness style for enhancing the quality of user selections, particularly among participants with lower domain familiarity. Moreover, allowing users to switch between conversational styles enables them to optimize their interaction based on personal preferences or situational needs, further improving task performance.

The main contributions of this work are as follows:
(1) We design and conduct a comprehensive user study to empirically investigate how different conversational styles impact preference elicitation, task performance, and user satisfaction in CRSs (Section~\ref{sec:experiment}).
(2) We develop a novel conversational scholarly assistant, specifically designed for this study, which are open-sourced to enable further research and development in the field of conversational recommender systems (Section~\ref{sec:csa}).
(3) We perform a quantitative analysis of the collected data, showing that different conversational styles are more effective depending on the level of user familiarity with the task domain (Section~\ref{sec:res}).
The code and study materials developed within this paper are made publicly available at \url{https://github.com/iai-group/umap2025-convstyles}.

%% file: umap2025-styles-02.tex
\section{Related Work}
\label{sec:related}

We are inspired by, build on, and position our work and contributions in the following realms of existing literature: conversational recommender systems (Section~\ref{sec:related:crs}), preference elicitation in CRSs (Section~\ref{sec:related:pe}), and conversational styles (Section~\ref{sec:related:cs}).

\subsection{Conversational Recommender Systems}
\label{sec:related:crs}

Early recommender systems, such as collaborative filtering~\citep{Sarwar:2001:WWW}, logistic regression~\citep{Nelder:1972:RSS}, and gradient boosting decision trees~\citep{Blanco:2013:ISWC}, rely heavily on historical data to predict user preferences.  While effective, these approaches struggle in cold-start scenarios where user data is sparse and fail to adapt to evolving preferences in real time. Additionally, they treat recommendation as a one-shot process, making them less suitable for high-involvement products (i.e., products bought rarely but chosen with more care and time) that require deliberation~\citep{Lee:2019:KDD, Jannach:2022:ACM}.

A \emph{conversational recommender system} (CRS) engages users in multi-turn conversations to elicit detailed, real-time preferences using natural language~\citep{Jannach:2022:ACM}. Beyond capturing explicit preferences, CRS can process feedback and explain recommendations.
\citet{Gao:2021:AI} identified five primary challenges in CRS research: effective preference elicitation, multi-turn strategy optimization, improving natural language understanding and generation, balancing exploration-exploitation trade-offs, and robust evaluation including user simulation.
This work shifts the focus of question-based preference elicitation from \emph{what} to ask to \emph{how} to ask it. Specifically, we investigate how conversational style influences preference elicitation, task completion, and user satisfaction.
We conduct this study in the scholarly domain, which serves diverse information needs, involves non-sensitive data, and engages users with domain expertise~\citep{Balog:2020:arXiv}.

\subsection{Preference Elicitation}
\label{sec:related:pe}

Preference elicitation is a core challenge in CRSs, involving two key questions:  \emph{What to ask?} and \emph{How to adjust the recommendations based on user responses?}~\citep{Gao:2021:AI}.
There are two primary approaches: item-based and attribute-based elicitation.
Item-based methods directly query users about specific items, updating recommendation lists based on feedback~\citep{Christakopoulou:2016:KDD, Zou:2020:SIGIR, Sepliarskaia:2018:RecSys}. This shifts recommendation models from static historical data to more dynamic, interactive systems.
Attribute-based methods focus on user preferences for item attributes and include critiquing-based techniques, where users refine recommendations by providing feedback on specific attributes~\citep{Wu:2019:RecSys, Luo:2020:SIGIR}. Reinforcement learning-based approaches further optimize attribute selection by adapting to user preferences over multiple interactions~\citep{Sun:2018:SIGIR, Lei:2020:WSDM}.
\citet{Kostric:2024:ACM} argued that users with lower domain expertise may struggle to answer attribute-based and proposed usage-oriented questions as an alternative. Our work employs an attribute-based approach but focuses on \emph{how} questions are asked rather than \emph{what} is asked. We investigate how conversational style influences user responses and recommendation effectiveness, potentially enabling more personalized and efficient interactions.

Prior work has also explored guiding user preference expression through nudging techniques, where systems directly suggest or emphasize specific areas for further inquiry~\citep{Gohsen:2023:CHIIRa}. Such techniques include repetition or speech modifications to steer users toward particular content.  In contrast, our study does not aim to guide users but instead observes the effects of conversational style on preference elicitation.

\subsection{Conversational Styles}
\label{sec:related:cs}

Conversational styles play a crucial role in shaping lexical communication. Early work by \citet{lakoff1979stylistic} categorized them into four types based on relational dynamics between participants: (1) \emph{clarity}, an ideal mode of discourse; (2) \emph{distance}, a style that does not impose others; (3) \emph{deference}, a style giving options; and (4) \emph{camaraderie}, direct expression of desires. Expanding on this, \citet{tannen1987conversational, Tannen:2005:Book} analyzed social discourse and identified two broad styles: \emph{high involvement}, characterized by active engagement and assertiveness, and \emph{high considerateness}, which prioritizes politeness and listener comfort.

In a relevant study, \citet{kim2019} compared survey response data quality acquired from a web platform in comparison to a chatbot having either a \emph{casual} or a \emph{formal} conversational style. The chatbot using a \emph{casual} conversational style attempted to establish a relationship with users using linguistic features that overlap with both of Tannen's \emph{high-involvement} and \emph{high-considerateness} conversational styles. The chatbot using \emph{formal} style is akin to the \emph{clarity style} (i.e., showing no involvement and the least relationship with the user) as summarized by Lakoff~\citep{lakoff1979stylistic}. Additionally, \citet{shamekhi2016exploratory} found that users prefer agents mirroring their own style, a finding reinforced by studies of information-seeking conversations~\citet{thomas2018style} using the MISC dataset~\citep{thomas2017misc}. 

While prior CRS work has explored human-like dialogue generation~\cite{li2022customized}, our study systematically examines the impact of conversational style on preference elicitation and user experience.  We operationalize Tannen’s \emph{high-involvement} and \emph{high-considerateness} styles to investigate their effects on CRS interactions.

Traditionally, response styles in dialogue systems were implemented via templates~\citep{Qiu:2020:CHI}. Recently, a new approach has emerged where large language models (LLMs) are used to simulate personas~\citep{Sun:2024:CUI, Zhong:2020:EMNLP, AliAmerJidAlmahri:2019:Informatics}, where personas are loosely defined fictional characters with traits such as tone, backstory, and personality \citep{Sutcliffe:2023:arXiv}.
Our work differs by controlling not just tone but also interaction dynamics, including pacing and engagement level.

%% file: umap2025-styles-03.tex
\section{Study Design}
\label{sec:experiment}

In this section, we detail the experimental design for our user study, where our main objective is to explore how conversational styles affect the preference elicitation process in CRSs and shape task performance and user satisfaction. Specifically, we contrast two conversational styles, which are detailed in Section~\ref{sec:exp:styles}: \emph{high involvement}, a fast-paced, direct, and proactive style, where the system takes the initiative and frequently offers suggestions, and \emph{high considerateness}, a polite and accommodating style that prioritizes clarity and alignment with users, allowing them to guide the interaction while the system provides support and additional information as needed.
The task given to the participants is described in Section~\ref{sec:exp:task}. The evaluation, where we focus on both objective and subjective measures, is described in Section~\ref{sec:exp:eval}.
The experiments are carried out on a prototype CSA system, which is detailed in Section~\ref{sec:csa}.

\subsection{Conversational Styles: High Involvement and High Considerateness}
\label{sec:exp:styles}

\begin{table}[t]
\centering
\small
\caption{A comparison between high-involvement and high-considerateness conversational styles, adopted from~\citep{Qiu:2020:CHI}.}
\begin{tabular}{p{3cm} >{\centering\arraybackslash}p{2cm} >{\centering\arraybackslash}p{2cm}}
\toprule
\textbf{Criteria} & \textbf{High \newline involvement} & \textbf{High \newline considerateness} \\ 
\midrule
C1. Rate of speech & fast & slow \\
C2. Turn-taking & fast & slow \\
C3. Introduction of topics & w/o hesitation & w/ hesitation \\
C4. Use of syntax & simple & complex \\
C5. Directness of content & direct & indirect \\
C6. Utterance of questions & frequent & rare \\ 
\bottomrule
\end{tabular}
\label{tab:communication_comparison}
\end{table}

In our study, we operationalize two distinct conversational styles for the CRS: \emph{high involvement} (\inv{}) and \emph{high considerateness} (\con{}), drawing on the framework developed by \citet{Tannen:2005:Book}. These conversational styles, originally observed in human-human interactions, reflect varying social and cultural norms that shape communication patterns. They have been adapted to human-agent interactions, with specific criteria guiding conversational agent design to emulate these styles~\citep{Qiu:2020:CHI}.

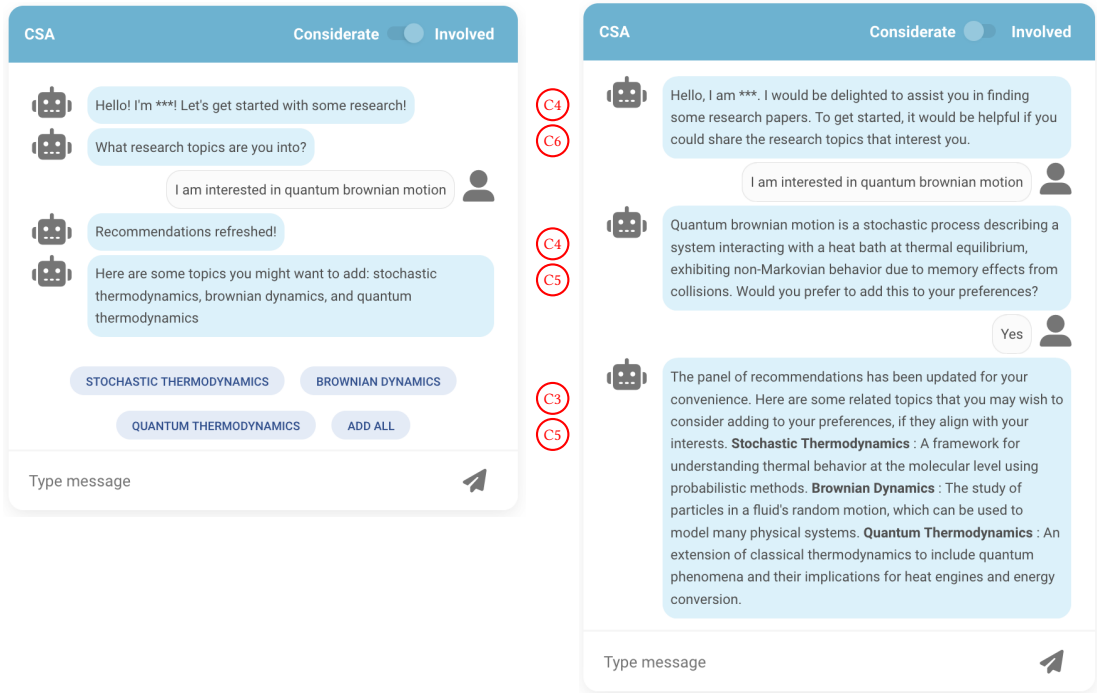
\begin{figure*}[t]
\centering
\begin{tabular}{p{0.41\textwidth} p{0.41\textwidth}}
\raisebox{.35\height}
{\begin{tikzpicture}
    \node[anchor=south west,inner sep=0] (image) at (0,0) {\includegraphics[width=0.39\textwidth]{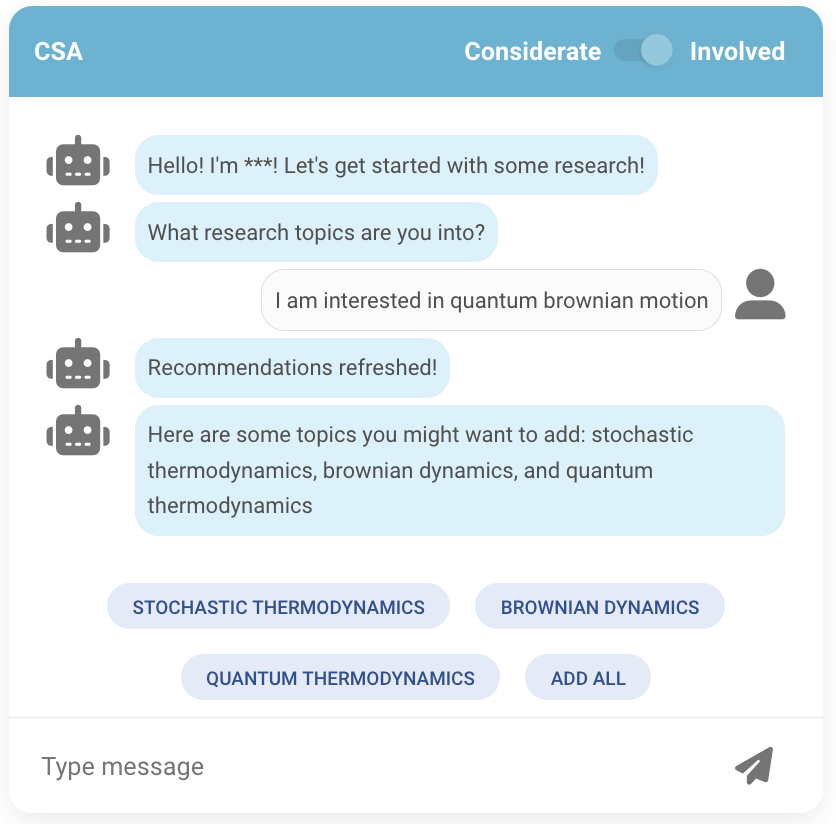}};
    \begin{scope}[x={(image.south east)},y={(image.north west)}]
        \draw[red, thick] (1.05, 0.8) circle[radius=0.03] node at (1.05, 0.8) {\scriptsize C4};
        \draw[red, thick] (1.05, 0.73) circle[radius=0.03] node at (1.05, 0.73) {\scriptsize C6};
        \draw[red, thick] (1.05, 0.53) circle[radius=0.03] node at (1.05, 0.53) {\scriptsize C4};
        \draw[red, thick] (1.05, 0.46) circle[radius=0.03] node at (1.05, 0.46) {\scriptsize C5};
        \draw[red, thick] (1.05, 0.23) circle[radius=0.03] node at (1.05, 0.23) {\scriptsize C3};
        \draw[red, thick] (1.05, 0.16) circle[radius=0.03] node at (1.05, 0.16) {\scriptsize C5};
    \end{scope}
\end{tikzpicture}} &
\begin{tikzpicture}
    \node[anchor=south west,inner sep=0] (image) at (0,0) {\includegraphics[width=0.39\textwidth]{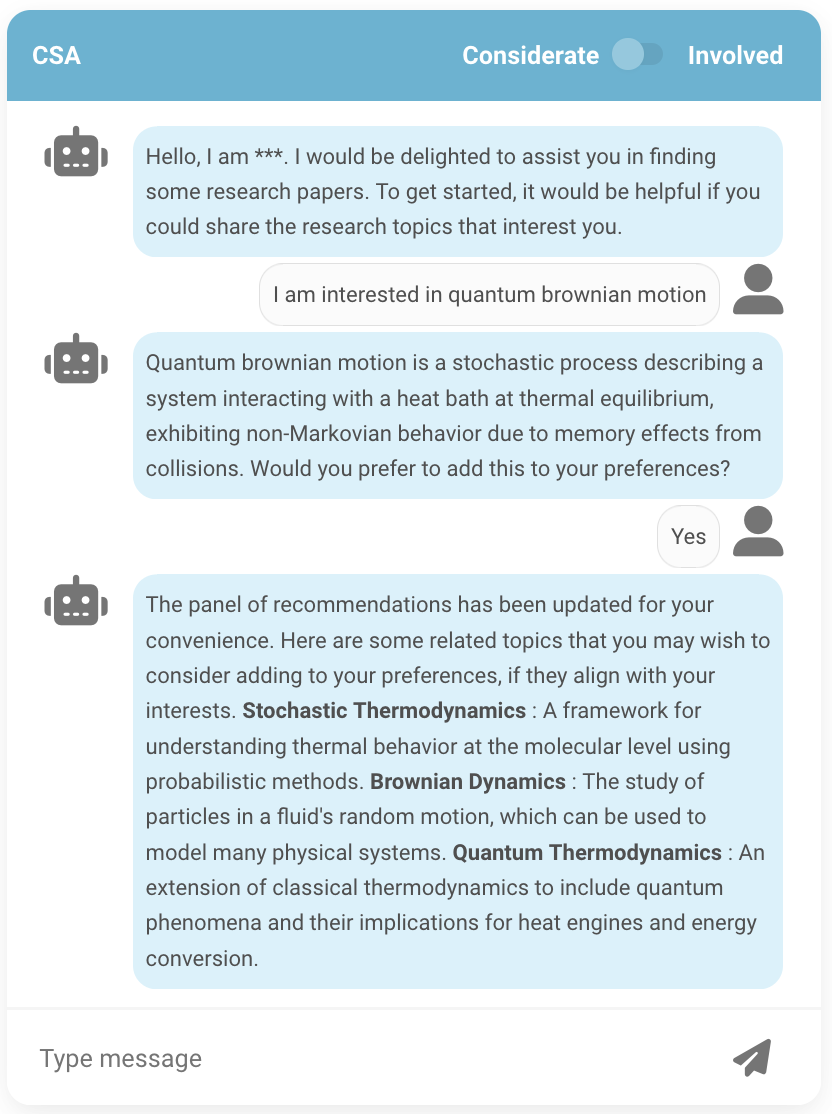}};
\end{tikzpicture}
\end{tabular}
\caption{Contrasting example conversations between two conversational styles for the same user need. \inv style on the left and \con style on the right.}
\label{fig:conversation_examples}
\vspace{-0.75\baselineskip}
\end{figure*}

Table~\ref{tab:communication_comparison} summarizes the key differences between the \inv{} and \con{} styles across six criteria (C1–C6). The criteria include \emph{rate of speech} (C1), \emph{turn-taking} (C2), \emph{introduction of topics} (C3), \emph{use of syntax} (C4), \emph{directness of content} (C5), and \emph{utterance of questions} (C6). In the \inv{} style, the system communicates with a fast rate of speech and rapid turn-taking, introduces topics without hesitation, uses simple syntax, is direct in content delivery, and asks questions frequently. Conversely, the \con{} style features a slower rate of speech and turn-taking, introduces topics with hesitation, employs complex syntax, communicates indirectly, and rarely asks questions.
Figure~\ref{fig:conversation_examples} shows excerpts from conversations contrasting the two styles where the user expresses the same information need in both cases.

To operationalize the criteria in our CRS, we group the criteria into two categories: \emph{Pacing and Tone} (C1, C2, C4, C5) and \emph{Content Introduction and Questioning} (C3, C6).

\subsubsection{\textbf{Pacing and Tone}} Pacing refers to how quickly the conversation progresses, influenced by the \emph{rate of speech} (C1) and \emph{turn-taking} (C2). In the \inv{} style, we simulate a fast rate of speech by displaying messages word by word with minimal delays between words (ranging from 5 to 50 milliseconds between each word), creating a sense of urgency. Shorter texts are displayed with slightly longer delays to simulate user typing speeds, preventing messages from appearing too abruptly. For longer texts, the delay is reduced to avoid making users wait too long.
Turn-taking is rapid, with the system responding promptly after the user's input, maintaining a dynamic interaction flow. Additionally, responses in this style are separated into several shorter messages, which contributes further to the sense of urgency. In contrast, the \con{} style features a slower rate of speech, with messages displayed character by character in a single long message and with longer delays (7 to 70 milliseconds between each character), fostering a more relaxed interaction. This is in stark contrast to the \inv{} style, where messages are presented word by word. The delay difference is highly discernible: the word-based delay in the \inv{} style contributes to a sense of urgency and momentum. In contrast, the character-based delay in the \con{} style slows the interaction down, encouraging users to process information carefully before responding.
Turn-taking is deliberate, with pauses before the system responds, allowing users more time to process information. This delay is built-in to cater for the time taken to generate detailed explanations.

Tone encompasses the \emph{use of syntax} (C4) and \emph{directness of content} (C5), affecting the system's manner of communication. The \inv{} style uses simple syntax and direct language, conveying information concisely. This approach creates an assertive and efficient tone.
The \con{} style employs more complex syntax and indirect language, often using polite expressions and detailed explanations. This results in a considerate and supportive tone.

\subsubsection{\textbf{Content Introduction and Questioning}} \emph{Content introduction} (C3) pertains to how the system introduces new topics or suggestions. In the \inv{} style, the system introduces topics without hesitation, frequently suggesting new keywords or preferences. It actively guides the conversation by offering options for the user to consider, using buttons to encourage users to act. In the \con{} style, the system introduces topics with hesitation, providing explanations before making suggestions. It waits for the user's cues by asking for confirmation before adding a preference, allowing the user to steer the conversation.

\emph{Utterance of questions} (C6) relates to how often the system asks questions. The \inv{} style asks questions frequently, aiming to gather information rapidly and keep the conversation moving forward. These questions are often direct and require immediate responses. The \con{} style asks questions rarely, and when it does, they tend to be open-ended or for confirmation to make sure there is alignment between the user and the agent. This approach reduces pressure on the user and encourages reflective responses.

\subsection{Task Design}
\label{sec:exp:task}

\begin{figure}[t]
    \centering
    \includegraphics[width=\linewidth]{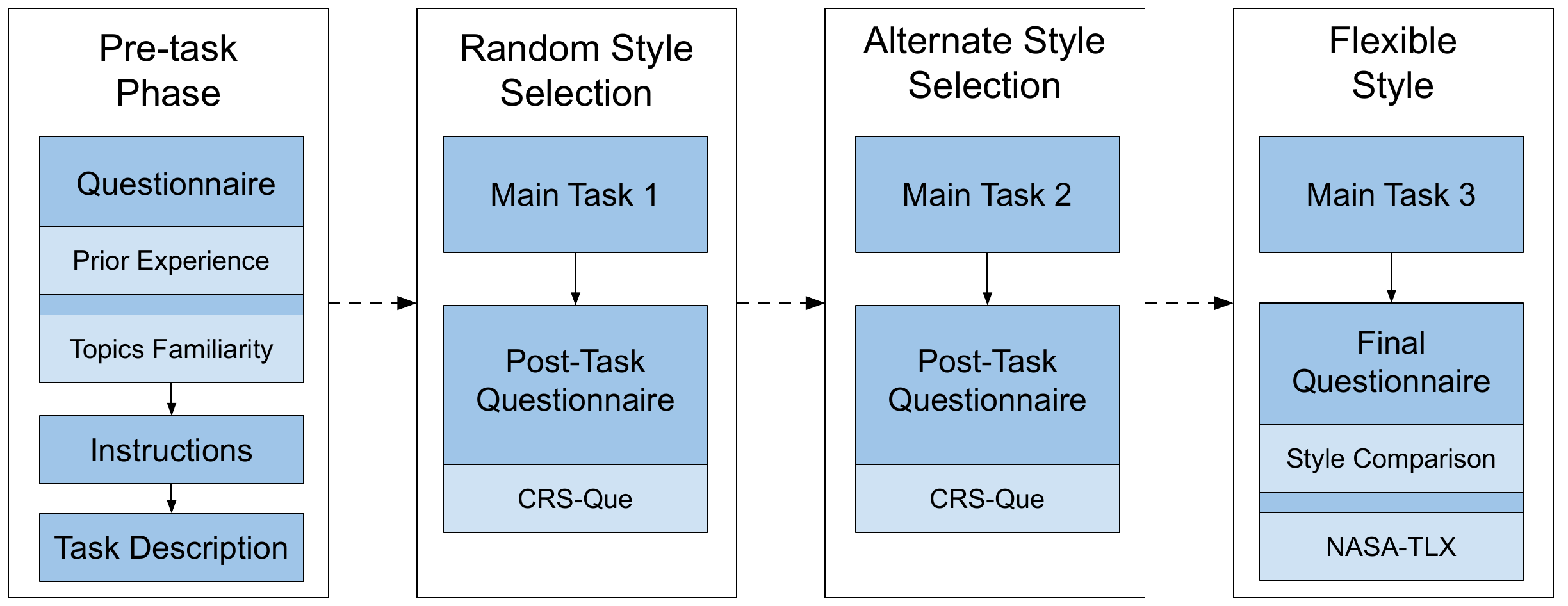}
    \caption{The user journey through the study phases: pre-task familiarization, main task interaction with the conversational system (counterbalanced with respect to which style is first encountered by participants), and post-task feedback. Participants completed three tasks in a randomized order with varying conversational styles, followed by feedback after each task.}
    \label{fig:journey}
    \vspace{-0.75\baselineskip}
\end{figure}

Figure~\ref{fig:journey} illustrates the user journey in our study, which consists of three experimental conditions with varying conversational styles of the CRS: \inv{} (high involvement), \con{} (high considerateness), and \agency{} (providing users with the agency to select and manipulate the conversational style of the CRS between \inv{} and \con{}). We employ a mixed experimental design, with conversational style as a within-subjects factor and the topic set as a between-subjects factor. Specifically, we recruit master's and Ph.D. students in computer science as participants to engage with each condition. We ask participants to use a scholarly conversational assistant to find research papers relevant to a given research topic.\footnote{In our study, a research topic is a specific subject or study area within a broader field.} To investigate the influence of prior knowledge, we divide participants into two groups. One group is presented with research topics from their primary domain (computer science). This allows us to assess how prior knowledge may affect their interaction with the CRS and task performance. In contrast, the second group engages with topics outside their primary domain of expertise (physics), ensuring their reliance on the CRS is not confounded by pre-existing knowledge. This approach allows us to isolate the CRS's performance from the participant's familiarity with the subject matter. To minimize order effects, the topic order is randomized for each participant.
The topics used in our study were carefully chosen based on the expected familiarity of our participants and the availability of relevant academic papers. The computer science topics were fictional and designed by the authors to reflect the knowledge students are expected to have in our university's applied data science curriculum at the time of the study. This design choice ensured the tasks were engaging and appropriately challenging for our participants. In contrast, the physics topics were actual master’s thesis topics provided to us by the Department of Physics at the University of Stavanger. To support the CRS functionality, we utilized papers from the arXiv\footnote{https://arxiv.org/} database. ArXiv was chosen for its extensive and up-to-date coverage of both Computer Science and Physics domains, making it an ideal source for scholarly literature recommendations.
The study is split into three phases: the pre-task phase, where participants familiarize themselves with the system; the main task phase, where they interact with the CRS to complete their assignment; and the post-task phase, where they provide feedback on their experience.

In the pre-task phase, participants begin by completing a questionnaire designed to capture their familiarity with conversational assistants and assess their prior knowledge of the research topics included in the study. The questionnaire includes questions on participants’ experience with conversational systems, such as how frequently they use these systems, their satisfaction with past interactions, and specific likes or dislikes. Additionally, they are asked to self-report their familiarity with two domains: Machine Learning and Neural Networks, as well as Theoretical Physics and Quantum Mechanics. This helps us gauge the participants’ perceived domain expertise before they engage with the system. Following the questionnaire, participants are provided with detailed instructions about the study, including screenshots of the system’s interface that highlight key functionalities. No formal training or hands-on exploration of the system is provided, as the design prioritizes an intuitive, conversational interaction experience.

During the main task, each participant engages with all three experimental conditions (\inv{}, \con{}, \agency{}) in a counterbalanced sequence. Simple randomization is employed for the first two conditions (\inv{} and \con{}) to control for order and learning effects, minimizing potential biases. This setup ensures an even distribution, with half of the participants beginning with the \inv{} condition and the other half starting with the \con{} condition. The \agency{} condition is consistently positioned as the final condition for all participants. This design choice was made to increase the ecological validity of users’ agency in selecting their preferred conversational style, based on an informed understanding and expectation obtained after exposure to both styles. We recruited 30 participants, all master’s or PhD students in the field of Computer Science, prioritizing academic expertise and domain knowledge as the primary inclusion criteria. General demographic information, such as age or gender, was not explicitly collected.

Participants in the study are asked to assume the role of a student, tasked with compiling a set of 5 research papers relevant to a given topic,\footnote{Participants in our study had no prior knowledge of the specific topics.} thereby simulating the process of gathering related work for a master thesis. 
The requirement to select 5 papers ensures sustained engagement with the conversational system and provides a clear objective. While interacting with the agent, participants bookmark relevant articles in real time to indicate their choices. This serves a dual purpose: it provides an end goal that participants work towards and allows us to gather data to assess their performance across the experimental conditions. While there is no algorithmic difference in paper recommendations between conditions, the interaction patterns and participant keyword choices influence the outcomes.

After completing each task, participants are asked to complete the CRS-QUE questionnaire~\citep{Jin:2024:ACM}, designed to evaluate user experience across four categories: Perceived Qualities, User Beliefs, User Attitudes, and Behavioral Intentions. CRS-QUE, which builds upon the ResQue~\citep{Pu:2011:RecSys} framework for recommender systems, captures both recommendation and conversational aspects of the system’s performance. 
Following the final condition (\agency), where participants can freely switch between the two conversational styles, they answer a set of custom questions. These questions focus on their preferred conversational style, the specific aspects they found appealing, and whether they interacted more frequently with one style over the other. Additionally, participants are asked to reflect on how their interaction might change based on familiarity with the research topics. Finally, the NASA-TLX~\citep{Hart:1988:Book} questionnaire is leveraged at this stage to assess the participants’ perceived cognitive load during their interactions with the system.

\subsection{Evaluation}
\label{sec:exp:eval}

The evaluation of our user study is two-fold, encompassing both subjective user experience and objective performance metrics. 
For the subjective component, we gather feedback using post-task questionnaires designed to capture participants’ perceptions of our CRS. Specifically, we employed the CRS-Que~\citep{Jin:2024:ACM} questionnaire, which asks participants to rate their satisfaction across several dimensions: perceived usefulness and ease of use (\emph{Perceived Qualities}), trust in the CRS (\emph{User Beliefs}), overall attitudes toward its use (\emph{User Attitudes}), and future usage intent (\emph{Behavioral Intention}). The responses are captured on a 7-point Likert scale, allowing us to quantify participants' overall satisfaction and engagement with the system.
Following the final task, open-ended questions are included to allow participants to elaborate on their experiences, offering qualitative insights into how the different conversational styles affected their interaction with the CRS.

In assessing the objective measures in our study, we focus on the efficiency of the conversational assistant. We do this by analyzing the number of interaction turns and time required for participants to identify relevant articles, both of which are standard metrics for evaluating the efficiency of task-oriented dialogue systems~\citep{Deriu:2021:AIR}. These metrics help us identify which conversational style leads to quicker and more effective recommendations and how this correlates with participants' subjective experiences.
To assess task performance, we evaluate the quality of the bookmarks participants collected during their interactions with the system. Relevance scoring for these bookmarked articles is obtained using a Large Language Model (LLM), building on the approach by \citet{faggioli2023perspectives}. Then, an expert in Computer Science verified the LLM-generated relevance scores, achieving substantial inter-rater agreement ($\kappa =0.78$).
Each article is assigned a relevance score of \textbf{0} (\textit{irrelevant}), \textbf{1} (\textit{somewhat relevant}), or \textbf{2} (\textit{highly relevant}), allowing us to determine how effectively the CRS helped participants identify the most relevant papers for their tasks. 

%% file: umap2025-styles-04.tex
\section{Conversational Scholarly Assistant System}
\label{sec:csa}

In this section, we describe the experimental system, i.e., the Conversational Scholarly Assistant (CSA), that we developed and used to conduct our study. In Section~\ref{sec:method:flow}, we describe the two conversational flows that users experienced, while in Section~\ref{sec:method:arch}, the system design and architecture are described.

\subsection{Conversational Flow}
\label{sec:method:flow}

\begin{figure}
    \centering
    \includegraphics[width=\linewidth]{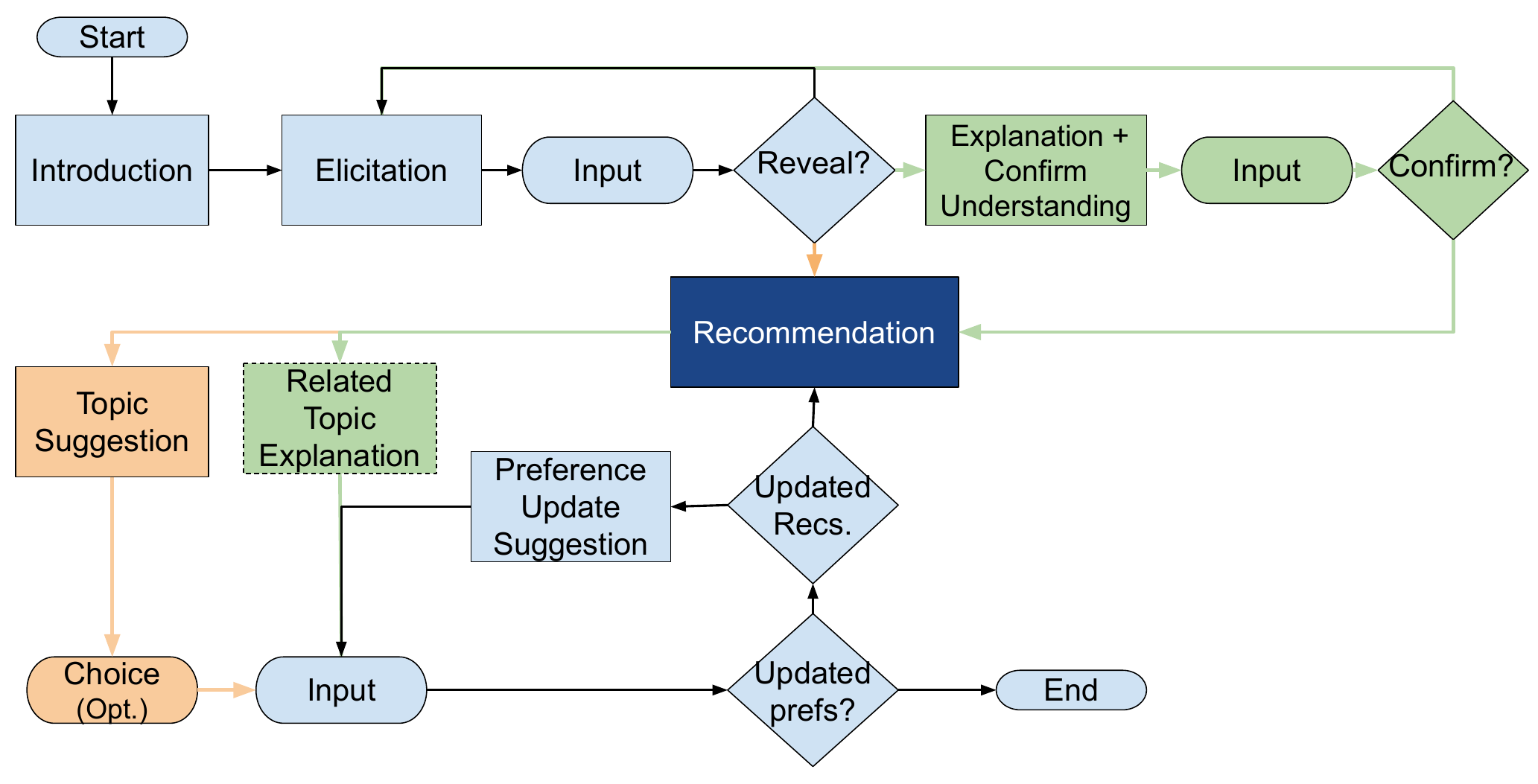}
    \caption{Dialogue flow. The \textcolor[HTML]{b6d7a8}{green} path shows the main flow for the \inv, while the \textcolor[HTML]{f9cb9c}{orange} shows the path for the \con conversational style.}
    \label{fig:flow}
\end{figure}

To address our research questions, we designed a system with two conversational flows characterized by distinct conversational styles. 
While the underlying task of eliciting preferences and presenting recommendations remains consistent across both styles, the interaction patterns differ to reflect the respective conversational styles. Figure~\ref{fig:flow} illustrates the core conversational flow, highlighting the differences between the \inv{} and \con{} styles. There are two main points of divergence in the flow.
First, after the initial elicitation and revealing of preferences, the \inv{} style proceeds directly to presenting recommendations. In contrast, the \con{} style engages in an additional round of interaction, providing explanations of keywords the user mentioned and confirming understanding before offering recommendations. This ensures that the user’s preferences are well-understood and aligned with the recommendations. 
The second distinction in the flow between the two conversational styles appears after recommendations are presented. In the \inv{} style, the system actively suggests related topics and encourages users to make choices via buttons (though free-text input is always available). This proactive approach aligns with the \inv{} style’s emphasis on taking initiative and maintaining a dynamic interaction flow. Conversely, in the \con{} style, the system provides explanations of related topics without nudging the user to select them, allowing users to steer the conversation according to their preferences.

Throughout both conversational flows, syntax, and tone are carefully chosen to match the intended style. The \inv style employs concise language and straightforward content, while the \con style opts for detailed explanations and a relatively more nuanced approach to information delivery (see Figure~\ref{fig:conversation_examples}).
 
Although there are notable differences in the operationalization of the two conversational styles, it is important to note that the functionality of the system remains fundamentally the same across these two styles. The contrasting conversational styles merely offer flexibility in how information is requested or presented. For instance, although the \inv style proactively makes suggestions without prompt, users can similarly request suggestions in the \con style. While the former may leverage buttons for quick responses, it also supports free text input, mirroring the latter's preference for open-ended dialogue. Furthermore, despite the frequent explanations characteristic of the \con style, users engaged in a \inv style conversation can still request detailed explanations whenever necessary.

\subsection{System Architecture and Modeling}
\label{sec:method:arch}

\begin{figure}
    \centering
    \includegraphics[width=\linewidth]{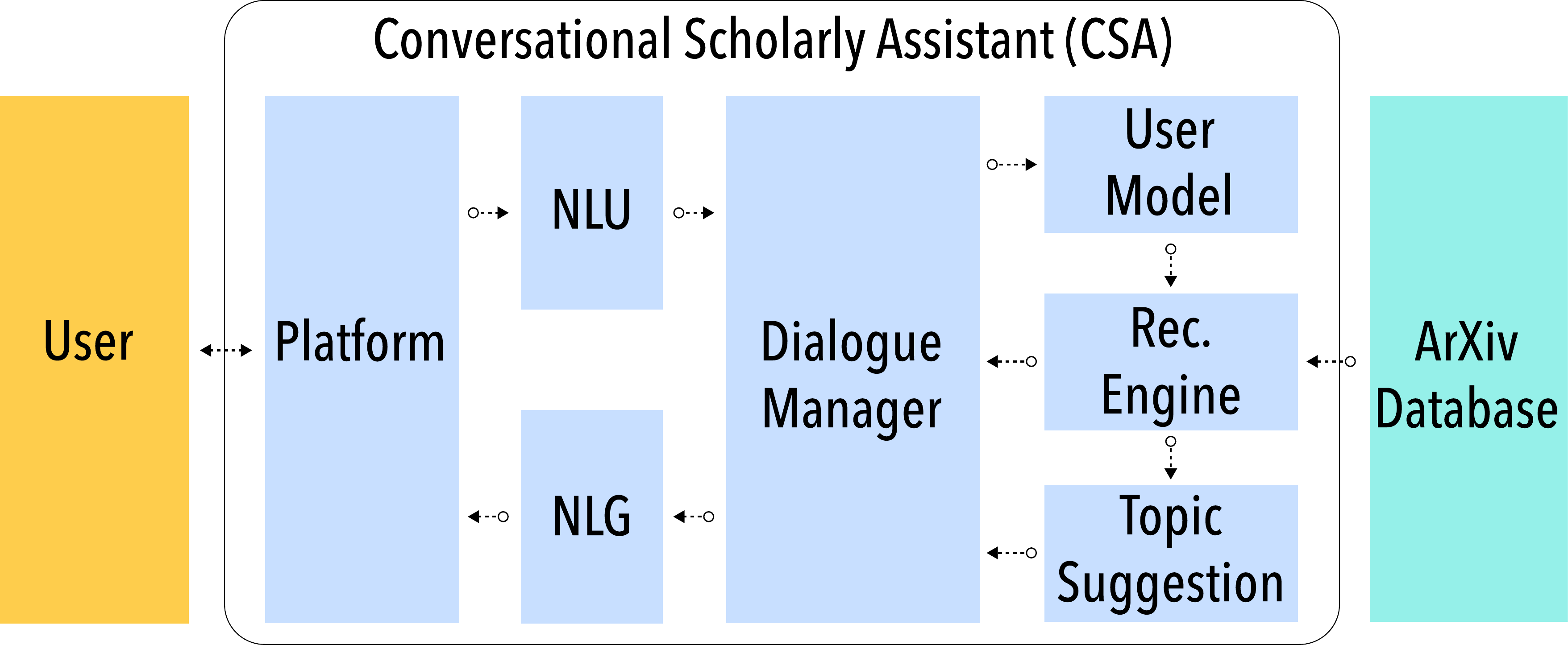}
    \caption{Architecture of the Conversational Scholarly Assistant (CSA) system used in this study.}
    \label{fig:architecture}
    \vspace{-0.75\baselineskip}
\end{figure}

\begin{table*}[t]
\centering
\small
\caption{Interaction metrics by conversational style. Note that for the number of turns and duration, lower is better. For elicited preferences and seen recommendations, higher is better. We report $\mu \pm \sigma$ (mean and standard deviation) and mixed-design ANOVA (style: within-subjects; domain: between-subjects) F-statistics with associated $p$-values. 
}
\label{tab:interaction_metrics}
\begin{tabular}{l|ccc|ccc}
\toprule
\textbf{Metric} & \textbf{\con} & \textbf{\inv} & \textbf{\agency} & \textbf{Style Effect} & \textbf{Domain Effect} & \textbf{Interaction Effect} \\
\midrule
Num. Turns & $16.79 \pm 11.15$ & $14.75 \pm 8.46$ & \textbf{\boldmath$14.08 \pm 7.7$} & $F(1, 22) = 0.42$, $p = .519$ & $F(1, 22) = 1.23$, $p = .277$ & \textbf{\boldmath$F(1, 22) = 4.49$, $p =  .045$} \\
Duration (min) & $13.17 \pm 6.19$ & \textbf{\boldmath$12.77 \pm 8.52$} & $13.54 \pm 9.84$ & $F(1, 22) = 0.38$, $p = .540$ & $F(1, 22) = 0.05$, $p = .816$ & $F(1, 22) = 0.00$, $p = .949$ \\
Recs. Seen & $20.58 \pm 12.91$ & \textbf{\boldmath$30.25 \pm 20.46$} & $21.92 \pm 11.65$ & \textbf{$F(1, 22) = 4.14$, $p = .054$} & \textbf{\boldmath$F(1, 22) = 11.23$, $p = .002$} & \textbf{\boldmath$F(1, 22) = 14.49$, $p = .000$} \\
Prefs. Elicited & \textbf{\boldmath$6.58 \pm 4.72$} & \textbf{\boldmath$6.58 \pm 6.20$} & $5.58 \pm 3.75$ & $F(1, 17) = 0.80$, $p = .381$ & $F(1, 17) = 0.00$, $p = .999$ & \textbf{\boldmath$F(1, 17) = 9.19$, $p = .007$} \\
\bottomrule
\end{tabular}
\end{table*}

Our system adopts a modular, component-based architecture~\citep[Chapter 15]{Jurafsky:2020:Book}.
This widely adopted architecture~\citep{Kostric:2022:RecSys, Bernard:2024:WSDM} comprises of 
three main modules: Natural Language Understanding (NLU), Dialogue Management (DM), and Natural Language Generation (NLG) (Figure \ref{fig:architecture}). Additionally, CSA integrates User Modeling, Recommendation, and Topic Suggestion modules to enhance personalization and relevance in the user experience.

The NLU module interprets user inputs by identifying intents and extracting relevant entities. We utilize a fine-tuned JointBERT model~\citep{Chen:2019:arXiv} based on SciBERT~\citep{Beltagy:2019:IJCNLP} for improved scholarly-domain understanding. The model recognizes a set of predefined user intents and actions, enabling the handling of various user requests. 
The DM module maintains the dialogue state and directs the conversational flow based on Section~\ref{sec:method:flow}. In the \inv{} style, the DM proactively offers suggestions and encourages rapid interactions, while in the \con{} style, it provides explanations and seeks confirmations to ensure alignment with the user’s intent.
The User Modeling module tracks and updates user preferences as keyphrases reflecting areas of interest or disinterest. The system is stateful, maintaining a persistent record of all user preferences throughout the conversation unless the user chooses to remove them. This allows the system to personalize the interaction by dynamically updating user profiles, ensuring more relevant recommendations. The Recommendation module retrieves research papers from an arXiv-based external database and ranks them using BM25. 
It first selects the top 100 papers based on positive preference keyphrases, then refines rankings by subtracting BM25 scores for negative preference keyphrases. This ensures papers aligned with user disinterests are ranked lower. The final top 10 papers are presented as recommendations.

The Topic Suggestion module analyzes recommended articles to extract the three most common keyphrases, leveraging the JointBERT model from the NLU module for consistency across system components. This enables dynamic topic suggestions closely aligned with the user’s current preferences.
The NLG module generates system responses using a hybrid approach: predefined templates for standard interactions and LLM-based dynamic generation for complex tasks. For most intents---such as eliciting preferences, acknowledging updates, or suggesting topics---the system selects from multiple templates per style to ensure variety and coherence. These templates are designed to align with the conversational tone, concise and direct for the \inv{} style, and more elaborate and polite for the \con{} style. For tasks requiring deeper explanation or justification, the system dynamically generates responses using an LLM. Our implementation employs LLama 3.1~\citep{Dubey:2024:arXiv} to provide rich, context-aware explanations.

%% file: umap2025-styles-05.tex
\section{Results and Analysis}
\label{sec:res}

Next, we present the findings of our study examining how conversational styles influence the usage of a CRS in the context of identifying relevant scientific literature on a given topic. We address the research questions by analyzing the impact of conversational styles on preference elicitation, task performance, and user engagement. We conducted ANOVA analyses to assess differences across conditions, applying a standard significance threshold of p < 0.05. 
Given the exploratory nature of our study and a small sample size, our primary objective is to observe trends and patterns rather than establish statistically significant effects.

\subsection{Effect on Preference Elicitation}
\label{sec:res:elicitation}

To investigate how conversational styles affect preference elicitation (\textbf{RQ1}), we analyzed interaction metrics across the three conversational styles: \con, \inv, and \agency. All 30 participants, master's and PhD students in computer science, completed tasks using each style within their randomly assigned domain—either \emph{Computer Science} (high familiarity) or \emph{Physics} (low familiarity). The average self-reported familiarity scores were significantly different ($p<0.001$), $5.2$ for Computer Science and $2.64$ for Physics on a 7-point Likert scale.

Table~\ref{tab:interaction_metrics} summarizes the interaction metrics. 
Although the average number of elicited preferences $(6.58)$ is identical for \con and \inv conditions and slightly higher than in the \agency condition, a significant style-by-domain interaction $(p=0.007)$ indicates that the effectiveness of these styles depends strongly on domain familiarity. 

For recommendations seen, participants in \inv viewed substantially more items $(30.25)$ than in \con $(20.58)$ or \agency $(21.92)$. While the style effect approaches significance $(p=0.054)$, domain familiarity significantly influences the number of items participants choose to view $(p=0.002)$, with a highly significant style-by-domain interaction $(p<0.001)$.
Additionally, participants in the \con condition spent more time interacting with the system (average duration of $13.17$ minutes) and engaged in more conversational turns (average of $16.79$ turns) compared to the \inv condition ($12.77$ minutes and $14.75$ turns). 
The number of turns shows a significant interaction effect $(p=0.045)$, indicating that the conversational style influenced interaction length differently across domains. The duration metric, however, did not yield significant effects.

Overall, these findings suggest that the \inv style is better suited for exploration, yielding more recommendations overall. The \con style, while resulting in a slightly lower number of preferences and unique recommendations, can enhance the quality of the preference elicitation by supporting users in refining their preferences through explanations and more thoughtful interactions.

\subsection{Effect on Task Performance}
\label{sec:res:performance}

To assess the impact of conversational styles on task performance (\textbf{RQ2}), we evaluated the quality of the bookmarks participants made during each task using two measures: (1) the average relevance scores assigned by a language model (LLM) used as an auto-rater and (2) the distribution of bookmarks across relevance levels. Each bookmarked article received a score of $2$ (highly relevant), $1$ (somewhat relevant), or $0$ (irrelevant), as described in the study design. 

\begin{table*}[t]
\centering
\small
\caption{Average relevance scores of bookmarks ($\mu \pm \sigma$: mean and standard deviation) with mixed-design ANOVA results (style: within-subjects; domain: between-subjects). 
}
\label{tab:quality_of_bookmarks_expanded}
\begin{tabular}{l|ccc|ccc}
\toprule
\textbf{Topic}  & \con & \inv & \agency & \textbf{Style Effect} & \textbf{Domain Effect} & \textbf{Interaction Effect} \\
\midrule
Computer Science  & $1.17 \pm 0.36$ & {\boldmath$1.30 \pm 0.57$} & $1.12 \pm 0.80$ & --- & --- & --- \\
Physics       & $1.15 \pm 0.70$ & $0.66 \pm 0.51$ & \textbf{\boldmath$1.25 \pm 0.64$} & --- & --- & --- \\
\midrule
\textbf{Overall}   & $1.16 \pm 0.56$ & $0.95 \pm 0.62$ & \textbf{\boldmath$1.19 \pm 0.70$} & \textbf{$F(1, 22) = 2.68$, $p = .115$} & \textbf{$F(1, 22) = 3.40$, $p = .078$} & \textbf{\boldmath$F(1, 22) = 7.78$, $p = .010$} \\
\bottomrule
\end{tabular}
\end{table*}

Table~\ref{tab:quality_of_bookmarks_expanded} presents the average relevance scores by conversational style and topic. In the Physics domain (low familiarity), the \agency style led to highest-quality bookmarks, with an average relevance score of $1.25 \pm 0.64$, compared to the \con ($1.15 \pm 0.70$) and \inv ($0.66 \pm 0.51$) styles. This indicates that the additional explanations in the \con style aided participants in selecting more relevant articles when they were less familiar with the topic.

Conversely, in the Computer Science domain (high familiarity), the \inv style resulted in the highest relevance scores ($1.30 \pm 0.57$), slightly outperforming both \con ($1.17 \pm 0.36$) and \agency ($1.12 \pm 0.80$). Participants with high domain expertise likely benefited from the efficiency of the \inv style, enabling them to quickly identify relevant keywords and articles without needing extensive explanations.

Overall, the style effect and domain familiarity alone were not significant. However, there was a statistically significant style-by-domain interaction ($p = 0.010$), confirming that the effectiveness of the conversational styles on bookmark quality strongly depends on the domain familiarity of users. The superior overall performance of the \agency condition ($1.19 \pm 0.70$), despite fewer elicited preferences (see Table~\ref{tab:interaction_metrics}), also suggests that allowing users flexibility in choosing conversational styles enhances task outcomes by catering to their situational needs or preferred interaction style.

\subsection{Effect on User Engagement}
\label{sec:res:engagement}

To explore the effect of conversational styles on user engagement (\textbf{RQ3}), we analyze the satisfaction scores from the post-task questionnaires (CRS-QUE). Participants rated their satisfaction on a 7-point Likert scale (1 = very unsatisfied, 7 = very satisfied).

\begin{table}[t]
\centering
\caption{User satisfaction scores by conversational style and topic ($Mean \pm Standard Deviation$)}
\label{tab:satisfaction_scores_full}
\begin{tabular}{l|rr|r}
\toprule
\textbf{Measure}  & \textbf{\con} & \textbf{\inv} & \textbf{Average}\\
\midrule
\textbf{Computer Science} & $4.01 \pm 1.52$ & \textbf{\boldmath$4.43 \pm 1.03$}  & $4.22 \pm 1.29$ \\
\midrule
\hspace{0.2cm}Perceived Qualities  & $4.28 \pm 1.77$ & {\boldmath$4.68 \pm 1.35$} & $4.48 \pm 1.58$\\
\hspace{0.2cm}User Beliefs & $3.72 \pm 1.93$ & {\boldmath$4.01 \pm 1.71$}  & $3.87 \pm 1.82$ \\
\hspace{0.2cm}User Attitudes  & $3.92 \pm 2.19$ & \textbf{\boldmath$4.65 \pm 1.62$} & $4.29 \pm 1.94$\\
\hspace{0.2cm}Behavioral Intention  & $3.85 \pm 2.08$ & \textbf{\boldmath$4.38 \pm 1.56
$} & $4.12 \pm 1.82$\\
\midrule
\textbf{Physics} & \textbf{\boldmath$4.63 \pm 1.16$} & $4.21 \pm 1.34$ & $4.41 \pm 1.26$ \\
\midrule
\hspace{0.2cm}Perceived Qualities  & \textbf{\boldmath$4.77 \pm 1.61$} & $4.35 \pm 1.69$ & $4.55 \pm 1.67$\\
\hspace{0.2cm}User Beliefs  & \textbf{\boldmath$4.40 \pm 1.67$} & $4.08 \pm 1.65$ & $4.23 \pm 1.66$\\
\hspace{0.2cm}User Attitudes  & \textbf{\boldmath$5.00 \pm 1.55$} & $4.26 \pm 1.68$ & $4.61 \pm 1.65$\\
\hspace{0.2cm}Behavioral Intention  & \textbf{\boldmath$4.20 \pm 1.93$} & $3.88 \pm 1.93$ & $4.03 \pm 1.91$ \\
\midrule
\textbf{Overall Satisfaction}  & $4.34 \pm 1.35$ & \textbf{$4.30 \pm 1.20$} & $4.32 \pm 1.26$ \\
\bottomrule
\end{tabular}
\end{table}

Table~\ref{tab:satisfaction_scores_full} shows the user satisfaction scores by conversational style and domain. Overall satisfaction scores are very similar between the \con $4.34 \pm 1.35$ and \inv ($4.30 \pm 1.20$) conditions. However, there are notable differences when considering domain familiarity. In the Computer Science domain, participants reported a higher satisfaction score in the \inv condition ($4.43 \pm 1.03$) compared to the \con condition ($4.01 \pm 1.52$). Specifically, scores for all four sub-measures (\emph{Perceived Qualities}, \emph{User Beliefs}, \emph{User Attitudes}, and \emph{Behavioral Intention}) were higher in the \inv condition, suggesting users with higher topic familiarity preferred the direct and efficient interaction offered by the \inv style. 

Conversely, in the \textbf{Physics} domain, participants reported a higher satisfaction score in the \con condition ($4.63 \pm 1.16$) compared to the \inv condition ($4.21 \pm 1.34$). The \con style scored consistently higher across all four sub-measures, indicating that users benefited from the additional explanatory support provided by this style when interacting with less familiar content.

These findings indicate that the alignment between conversational style and user topic familiarity influences user satisfaction with the CRS. This supports our previous observation that users with higher familiarity found the \inv style more satisfying due to its efficiency, while those with lower familiarity preferred the \con style for its supportive explanations.

%% file: umap2025-styles-06.tex
\section{Conclusions and Future Work}
\label{sec:concl}

We investigated how conversational styles---high involvement and high considerateness---affect preference elicitation, task performance, and user satisfaction in a Conversational Recommender System (CRS). Through a controlled user study involving 30 master’s and PhD students, we found that conversational style influences user interaction and outcomes. The involved style elicited more user preferences and presented more recommendations, particularly benefiting users with high domain familiarity.  In contrast, the considerate style led to higher-quality bookmarks in low-familiarity domains, suggesting that a slower pace and detailed explanations help users select more relevant articles. 
Allowing users to switch between styles resulted in the highest task performance, indicating the potential for dynamically adapting interactions to suit individual user preferences and situational demands. 

However, in interpreting these results, it is important to consider certain limitations. The small sample size and focus on a single domain may limit the generalizability of our findings. Furthermore, the fixed ordering of the Agency condition may have introduced order effects, as participants could have been influenced by prior exposure to the other styles or experienced fatigue by the final task.
Expanding to larger participant pools and application domains will validate and extend these findings beyond the discussed limitations.

This study raises ethical considerations around user experience, privacy, and undue influence. Conversational styles not only shape how users engage with the system but also how they articulate and refine their preferences, which can affect decision-making. While some styles may be more effective for certain users, system-driven choices should not unintentionally nudge users toward specific preferences or limit their ability to explore alternatives. Additionally, as our study focuses on non-sensitive academic recommendations without collecting personal data, privacy risks are minimized.

In future work, we aim to explore automatic user expertise detection to enable CRS systems to dynamically tailor interactions. This could further enhance the flexibility and effectiveness of CRSs in real-time recommendation scenarios.